\documentclass[sigconf, 9pt]{acmart}

\renewcommand\footnotetextcopyrightpermission[1]{}

\AtBeginDocument{%
  }

\usepackage{amsmath}
\usepackage{algorithmic}
\usepackage{algorithm}
\usepackage{graphicx}
\usepackage{pifont}
\usepackage{textcomp}
\usepackage{xcolor}
\usepackage{booktabs} 
\usepackage{soul}
\graphicspath{{figs/}}
\usepackage{braket}
\usepackage{etoolbox}
\usepackage{fancyhdr}
\usepackage{multirow}
\usepackage{makecell}

\newcommand{\cmark}{\ding{51}} % Define \cmark for a checkmark
\newcommand{\xmark}{\ding{55}} % Define \xmark for a cross

\begin{document}

\fancyhead{}
\title{Adversarial Data Poisoning Attacks on Quantum Machine Learning in the NISQ Era}

\author{Satwik Kundu}
\affiliation{%
  \institution{Pennsylvania State University}
  \city{University Park}
  \state{PA}
  \country{USA}
  \postcode{16801}}
\email{satwik@psu.edu}

\author{Swaroop Ghosh}
\affiliation{%
  \institution{Pennsylvania State University}
  \city{University Park}
  \state{PA}
  \country{USA}
  \postcode{16801}}
\email{szg212@psu.edu}

% \renewcommand{\shortauthors}{Kundu et al.}

\begin{abstract}
With the growing interest in Quantum Machine Learning (QML) and the increasing availability of quantum computers through cloud providers, addressing the potential security risks associated with QML has become an urgent priority. One key concern in the QML domain is the threat of data poisoning attacks in the current quantum cloud setting. Adversarial access to training data could severely compromise the integrity and availability of QML models. Classical data poisoning techniques require significant knowledge and training to generate poisoned data, and lack noise resilience, making them ineffective for QML models in the Noisy Intermediate Scale Quantum (NISQ) era. In this work, we first propose a simple yet effective technique to measure intra-class encoder state similarity (ESS) by analyzing the outputs of encoding circuits. Leveraging this approach, we introduce a \underline{Qu}antum \underline{I}ndiscriminate \underline{D}ata Poisoning attack, QUID. Through extensive experiments conducted in both noiseless and noisy environments (e.g., IBM\_Brisbane's noise), across various architectures and datasets, QUID achieves up to $92\%$ accuracy degradation in model performance compared to baseline models and up to $75\%$ accuracy degradation compared to random label-flipping. We also tested QUID against state-of-the-art classical defenses, with accuracy degradation still exceeding $50\%$, demonstrating its effectiveness. This work represents the first attempt to reevaluate data poisoning attacks in the context of QML.
\end{abstract}

%    <concept>
%        <concept_id>10010520.10010521.10010542.10010550</concept_id>
%        <concept_desc>Computer systems organization~Quantum computing</concept_desc>
%        <concept_significance>500</concept_significance>
%        </concept>
%    <concept>
%        <concept_id>10010520.10010521.10010542.10010294</concept_id>
%        <concept_desc>Computer systems organization~Neural networks</concept_desc>
%        <concept_significance>500</concept_significance>
%        </concept>
%  </ccs2012>

% \ccsdesc[500]{Computer systems organization~Quantum computing}
% \ccsdesc[500]{Computer systems organization~Neural networks}
% \keywords{Quantum machine learning, noisy intermediate-scale quantum, data poisoning attack, quantum cloud security, noise resilience.}


\maketitle

\section{Introduction}
%\hl{Fixed!} 
Quantum computing is rapidly progressing, with companies like Atom Computing and IBM recently unveiling the largest quantum processors ever developed, boasting 1,225 and 1,121 qubits, respectively \cite{computing2023quantum, gambetta2023hardware}. The significant interest in quantum computing stems from its potential to offer substantial computational speedups over classical computers for certain problems. Researchers have already begun leveraging current noisy intermediate-scale quantum (NISQ) machines to demonstrate practical utility in this pre-fault-tolerant era \cite{kim2023evidence}. Within this emerging field, quantum machine learning (QML) has also gained considerable attention, merging the power of quantum computing with classical machine learning algorithms. QML explores the potential to improve learning algorithms by leveraging unique quantum properties like superposition and entanglement, opening new horizons in computational speed and capability. Several QML models have been explored, with quantum neural networks (QNNs) \cite{schuld2014quest, abbas2021power, schuld2020circuit} standing out as the most popular model, mirroring the structure and function of classical neural networks within a quantum framework.

However, the rapidly evolving field of quantum computing brings not only unique computational capabilities but also a variety of new security challenges. Several recent works have explored the security aspects of quantum circuits in current NISQ-era. 
%For instance, the authors in \cite{xu2024security} identified multiple vulnerabilities at the gate-to-pulse level interface of quantum circuits and proposed corresponding defense frameworks. Another study 
Authors in \cite{erata2024quantum} introduced a power side-channel attack on quantum computer controllers, demonstrating how power trace analysis could be used to reconstruct quantum circuits. Researchers have also begun analyzing the effectiveness of adversarial attacks on QML models, such as model stealing attacks \cite{kundu2024evaluating}, where an adversary can replicate the functionality of a black-box QNN using only input-output queries. Novel defenses have been proposed to mitigate such attack scenarios \cite{kundu2024stiq}. Additionally, recent works have highlighted the threat of backdoor attacks \cite{chu2023qdoor, chu2023qtrojan} on QNNs, revealing them to be a significant concern for QML models. Another critical threat to QML models in current cloud settings, considered the greatest concern for classical ML deployment \cite{grosse2023machine, kumar2020adversarial}, is data poisoning attacks, which have not yet been explored in the context of QML.

Data poisoning attacks, first proposed in \cite{biggio2012poisoning}, are a class of adversarial attacks aimed at manipulating a victim model by either injecting poisoned samples into the dataset or perturbing a subset of the training data $\mathcal{D}_{tr}$. A detailed description of data poisoning attacks and their sub-classes can be found in Section \ref{poisoning-attack}. These attacks have been extensively studied in the classical domain, where most approaches rely on a deep understanding of the victim's training procedure (unrealistic) and gradient optimization techniques to generate minimally perturbed poisoned samples (costly). Furthermore, such methods often lack robustness to the noise present in modern NISQ devices, highlighting the need for a fresh perspective on data poisoning attacks in QML. Building on how data poisoning initially focused predominantly on label-flipping attacks, we take a first step in exploring indiscriminate data poisoning in QML by proposing a novel label-flipping technique based on intra-class state similarity in quantum Hilbert space.

In this work, we propose a quantum indiscriminate data poisoning attack, QUID, using a novel label-flipping strategy. QUID assigns poisoned labels to a subset of the training data such that the encoded state is furthest from its true class. Unlike other classical label-flipping \cite{taheri2020defending, paudice2019label} that require separately training the victim model to find adversarial labels, QUID avoids this costly training process while being noise-robust, as it can effectively determine poisoned labels by executing the encoding circuit on noisy hardware. Our main contributions in this work are:
\begin{itemize} 
    \item Introduced intra-class encoder state similarity (ESS) in the quantum Hilbert space using density matrices derived solely from the encoding circuit. ESS can be extensively used to estimate the performance of state preparation circuits for any given dataset without the need to train QML models from scratch. 
    \item Leveraged ESS to propose a novel label-flipping attack, QUID, which determines optimal poisoned labels without requiring knowledge of the victim model's training procedure or the need for any additional training.
    \item Evaluated the performance of QUID against random flipping and randomly poisoned samples across various QNN architectures and datasets in both noiseless and noisy environments, along with testing against a state-of-the-art defense technique \cite{levine2020deep}, demonstrating its effectiveness.
\end{itemize}
\textit{To the best of our knowledge, this is the first work to evaluate data poisoning attacks in the context of QML.}
\section{Background } \label{background}

\subsection{Quantum Neural Network (QNN)} 
A QNN comprises three key components: (i) a classical-to-quantum data encoding circuit, (ii) a parameterized quantum circuit (PQC) optimized for specific tasks, and (iii) measurement operations. For continuous variables, angle encoding is widely used, mapping classical features to qubit rotations along a chosen axis \cite{abbas2021power}. Since qubit states repeat every \(2\pi\), features are typically scaled to \(0\) to \(2\pi\) (or \(-\pi\) to \(\pi\)) during preprocessing. In this study, we use \(RZ\) gates for encoding classical features into quantum states.

\begin{figure}[!t]
        \vspace{-4mm}
        \centering 
        \includegraphics[width=\linewidth]{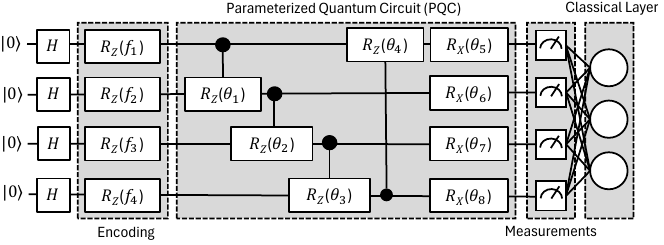}
        \vspace{-5mm}
        \caption{Architecture of a 4-qubit hybrid QNN. Classical features are encoded as angles of rotation gates (\( R_Z \)). The PQC transforms the encoded states to explore the search space and entangle features. Measured expectation values are then fed into a classical linear layer for the final prediction.}
        \label{qnn_circuit}
        \vspace{-4mm}
\end{figure}

A PQC consists of parameterized quantum gates designed to solve specific problems and serves as the only trainable block in a QNN for pattern recognition. It includes entangling operations to generate correlated multi-qubit states and parameterized single-qubit rotations for exploring the solution space. Measurement collapses qubit states to `0' or `1', and we use the expectation value of Pauli-Z to determine the average qubit states. These values are fed into a classical neuron layer (equal to the dataset's class count) for final classification in our hybrid QNN, as shown in Fig. \ref{qnn_circuit}. Other QML architectures may apply softmax directly or use additional classical layers for processing.

\subsection{Data Poisoning Attack} \label{poisoning-attack}
%\hl{Added!} 
Data poisoning attacks \cite{biggio2012poisoning, lu2022indiscriminate} are a class of adversarial attacks aimed at manipulating data by either tampering the current samples/labels or injecting poisoned samples to compromise the integrity and/or availability of the model. Data poisoning attacks can be broadly classified into three main categories: i) \textit{Targeted} attacks, which preserve the functionality and behavior of the system for general users while causing misclassification for specific target samples; ii) \textit{Indiscriminate} attacks, which aim to degrade model performance on test data; and iii) \textit{Backdoor} attacks, which, for any test sample containing a specific pattern, known as the backdoor trigger, induce misclassification without affecting the classification of clean test samples. In this work, we focus on indiscriminate data poisoning attacks on QNNs which can further be divided into; \textit{a) Label-flip} poisoning, where adversary does not perturb the features but only flips label of a subset of the dataset, \textit{b) Bilevel} poisoning, where attacker can obfuscate both the training samples and their labels and \textit{c) Clean-label} poisoning where attackers only slightly perturb each sample to preserve it's class i.e. performing a clean-label attack. Here, we propose a quantum state proximity based label-poisoning attack.

\begin{table}[!t]
    \vspace{-4mm}
    \centering
    \caption{Comparison with related classical attacks.}
    \label{tab_related_works}
    \vspace{-2mm}
    \begin{tabular}{cccccc}
    \cmidrule(lr){3-5}
    \multicolumn{1}{c}{} & \multicolumn{1}{c}{} &\multicolumn{3}{c}{Attacker's Knowledge} & \multicolumn{1}{c}{}\\
    \cmidrule(lr){1-6}
    Attacks    & \makecell{Noise \\ Resilient} &\makecell{Training \\ Data} & \makecell{ Training \\ Procedure} & \makecell{Test \\ Data} &  \makecell{Requires \\ Training} \\
    \cmidrule(lr){1-6}
    \cite{paudice2019label}              & \xmark  & \cmark          & \cmark      &  \xmark     & \cmark         \\
    \cite{koh2022stronger}       & \xmark  & \cmark          & \cmark      &  \cmark     & \cmark   \\
    % \cite{geiping2020witches}           & \xmark  & \cmark          & \cmark      &  \xmark     & \cmark     \\
    \cite{lu2023exploring}           & \xmark  & \cmark          & \cmark      &  \xmark     & \cmark     \\
    \textbf{Ours}           & \cmark  & \cmark          & \xmark      &  \xmark     & \xmark \\
    \bottomrule
    \end{tabular}
    \vspace{-4mm}
\end{table}

\subsection{Related Works and Limitations} \label{related_works}

%\hl{Added!} 
Although there has been work on backdoor attacks in QNNs \cite{chu2023qtrojan, chu2023qdoor}, some of which modify QNN architecture to perform indiscriminate attacks, no research has yet explored the viability of data poisoning attacks on QNNs. Current state-of-the-art (SOTA) classical indiscriminate data poisoning attacks \cite{lu2023exploring, lu2022indiscriminate} are gradient-based and assume that the adversary has access to the training data ($\mathcal{D}_{tr}$), the target model (white-box access), the training procedure (loss function, optimization process, hyperparameters, etc.), and sometimes even the test data \cite{koh2022stronger}. QUID, on the other hand, only requires access to $\mathcal{D}_{tr}$ and gray-box access to the QNN circuit, i.e., only the encoding circuit of the QNN. This is a more realistic and efficient framework considering the high cost associated with trainings QNN on NISQ devices. 

Apart from relying on extensive knowledge and expensive optimization procedures to poison the dataset, most classical SOTA attacks ensure that the poisoned samples, i.e., $\{x_i \mid (x_i, y_i) \in \mathcal{D}_{p} \subset \mathcal{D}_{tr}\}$, are minimally perturbed to avoid detection. For example, the Gradient Canceling (GC) attack \cite{lu2023exploring} aims to manipulate the model’s training process by injecting slightly perturbed poisoned data into $\mathcal{D}_{tr}$, with the goal of disrupting learning by ``canceling out'' the gradients that guide parameter updates during training. However, due to the noisy nature of current quantum hardware, these small perturbations, effective in classical deep learning, are masked by the inherent quantum device noise, rendering them ineffective (Table~\ref{tab_related_works}). Studies have even shown that inherent noise makes QNNs adversarially robust \cite{du2021quantum, wang2021roqnn}. Consequently, there is a need to reevaluate data poisoning attacks in the context of QML.

While label-flipping attacks exist in classical ML, QUID’s novelty lies in adapting this idea to QML setting by leveraging quantum properties, such as state similarity in Hilbert space. 
%Later in Section~\ref{related_works} and Table~\ref{tab_related_works}, we also discuss why classical techniques don’t directly translate to quantum models. 
To date, no prior work has focused on poisoning attacks or defenses specific to QML models, which is also why quantum-specific defenses remain unexplored. We acknowledge this limitation in Section~\ref{sec:results}.

% \hl{Idea?, types: Targeted, Indiscriminate \& Backdoor. Classical techniques: GC, GradPC and TGDA drawbacks for quantum?}

%\section{Related Works}
%Data theft poses a substantial security risk to emerging QML platforms particularly with the advent of third-party quantum cloud providers, where QML circuits submitted for training are highly vulnerable. To address the generic quantum circuit security risks, strategies have been explored, such as by authors in \cite{upadhyay2022robust}, which suggests dividing circuit execution between trusted and untrusted providers to mitigate adversarial interference risks. There have also been obfuscation based strategies to protect quantum circuits from less trustworthy cloud providers \cite{upadhyay2022robust, kundu2024evaluating} and compilers \cite{suresh2021short}. Recently there has also been works to protect VQAs and QMLs from untrusted cloud \cite{patel2023toward, ayanzadeh2023enigma, kundu2024stiq}. Despite these efforts, to the best of our knowledge, there have not yet been any methods developed that effectively protect the data encoded into QML circuits from cloud-based adversaries.

%\hl{QML security works.}

\section{Adversarial Framework}

\subsection{Attack Model}
%\hl{Fixed!} 
We consider a scenario where the victim sends both the pre-processed and normalized data, $\mathcal{D}_{tr}$, and the QNN ($\mathcal{F}$) to the quantum cloud provider for training. Although the current pipeline involves sequentially encoding data into the QNN one sample at a time and sending it to the cloud for execution, in the near future, a more efficient pipeline will likely be adopted to avoid network latency delays. 
%In this improved setup, the cloud provider would have access to both $\mathcal{D}_{tr}$ and $\mathcal{Q}$. 
However, we assume the presence of an adversary, unknown to the victim, within the quantum cloud. This adversary could be a malicious insider managing the quantum cloud, or an external attacker breaching its defenses.
%, or a vulnerability exposing the training data and quantum circuits stored on quantum cloud servers. 
Our primary concern is the integrity and the availability risk posed by these adversaries. Specifically, the adversary can modify $\mathcal{D}_{tr}$, to degrade the QNN's performance, making it ineffective and unusable for the victim. 

% We consider a general scenario in which pre-processed and normalized data, $X$, is encoded into a QML circuit for execution on quantum hardware via a cloud service, either for training or inference. However, we assume that the quantum cloud, where the QNN circuit is executed, is untrusted \hl{and unknown--why?} to the QNN owner \hl{mention insider adversary in otherwise trusted quantum cloud as well}. This lack of trust could stem from several potential threats, such as malicious insiders managing the quantum cloud, external attackers breaching its defenses, or vulnerabilities that expose queued quantum circuits. Our primary concern is the confidentiality risks posed by these adversaries. Specifically, since the cloud has full access to the circuit, a critical threat is the stealth of the data encoding component of the QML model. If an adversary is able to sequentially track the circuits sent for execution and their corresponding data encoding components, they can steal the embeddings, posing a serious threat to the model's intellectual property (IP). This is because, acquiring high-quality data is particularly challenging in specialized domains, requiring significant effort, time, and adherence to regulatory standards. Additionally, novel techniques are often used to preprocess raw data into a reduced latent space that can be efficiently encoded into QML models, ensuring optimal performance.v

\begin{figure}[!t]
        \vspace{-4mm}
        \centering 
        \includegraphics[width=0.9\linewidth]{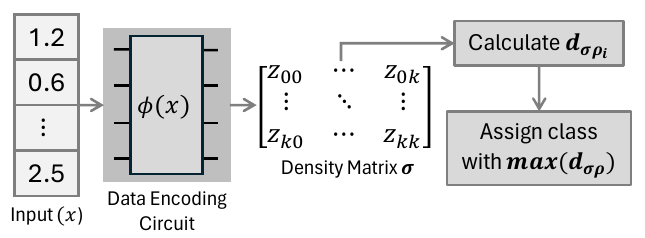}
        \vspace{-2mm}
        \caption{Overview of QUID's label-poisoning technique. The adversary extracts the encoding circuit from the QNN and uses it to compute the corresponding output density matrix ($\sigma$) for a portion ($\varepsilon$) of the training data $\mathcal{D}_{train}$. It then calculates the matrix distance between $\sigma$ and the remaining samples in the dataset ($\rho_i$), assigning the class with the maximum distance ($\max d_{\sigma\rho}$).}
        \label{fig:idea_overview}
        \vspace{-2mm}
\end{figure}

\subsection{Adversary Objective}
%\hl{Fixed!} 
The primary goal of the adversary is to poison a portion ($\varepsilon$) of the training dataset ($\mathcal{D}_{tr}$) in a way that maximally degrades the QNN's performance on the entire test set ($\mathcal{D}_{test}$). The adversary may be motivated by various factors, such as gaining a competitive advantage in the market or deceiving the victim into altering the QNN training specifications (e.g., PQC architecture, or hyperparameters). Such actions could require either retraining the QNN or training for more epochs, ultimately resulting in greater profit for the cloud-based adversary. Here, we consider that the adversary only poisons a portion ($\varepsilon$) of $\mathcal{D}_{tr}$ and does not inject additional poisoned samples to it. Thus, total size of the training dataset i.e. $|\mathcal{D}_{tr}|$ is fixed.
%\hl{can poison xxx complete the sentence.}

%The primary goal of the adversary is to steal the data encoding circuit to avoid paying for the highly sensitive training or inference data embedded in it. By obtaining the circuit, the adversary could sell it for profit or leak it to competitors, posing a threat to intellectual property and allowing others to gain insights into the victim's business processes or technological innovations. Additionally, the adversary could use the stolen circuit to train an QML model and offer a paid query-based service, similar to Machine Learning as a Service (MLaaS).

\subsection{Adversary Knowledge}
%\hl{Fixed!} 
We assume that the cloud-based adversary has access to the training data, $\mathcal{D}_{tr} = {(x_i, y_i)}_{i=1}^n$, and gray-box access to the QNN circuit ($\mathcal{F}$) executed on the quantum hardware. This means the adversary has knowledge of the data encoding circuit of the QNN but does not necessarily have access to the rest of the circuit or the training procedure. This is a reasonable assumption since the victim needs to define how the training data is encoded into the QNN, thereby giving the adversary direct access to the state preparation (encoding) circuit. %\hl{why not white box access and only grey box?--justify}

%that the adversary has full knowledge of the circuit architecture, parameters, and the various components of the QML model, including the data encoding circuit, PQC, and the measurement operation. Additionally, we assume the adversary can distinguish the data encoding circuit from the rest of the QML circuit but cannot extract the classical data embedded in it. These assumptions are reasonable for two main reasons. 

%First, since the parameters of the PQC remain constant during the execution of a batch of samples, while the parameters of the data encoding circuit change with each sample, the adversary can identify the data encoding circuit by observing which parameters vary. Second, due to the fact that there is a wide variety of embedding techniques, such as amplitude encoding and angle encoding, which make it extremely challenging for the adversary to determine the encoded classical data by simply analyzing the circuit architecture and parameters, especially after the circuit has been transpiled. 

%As a result, given that the adversary has access to the circuit and it's parameters, it has the ability to easily extract the classical data encoded into it. which inturn poses a significant risk to the not only the whole data encoding circuit but also the classical data embedded in the circuit.

\section{Proposed Methodology}
Fig.~\ref{fig:idea_overview} provides a high-level overview of our poisoning technique. Before delving into the details, we first define the notion of Intra-Class Encoder State Similarity (ESS), which is leveraged to poison the training data $\mathcal{D}_{tr}$. Subsequently, we outline the label-poisoning strategy employed by QUID.

\begin{figure}[!t]
        \vspace{-4mm}
        \centering 
        \includegraphics[width=0.9\linewidth]{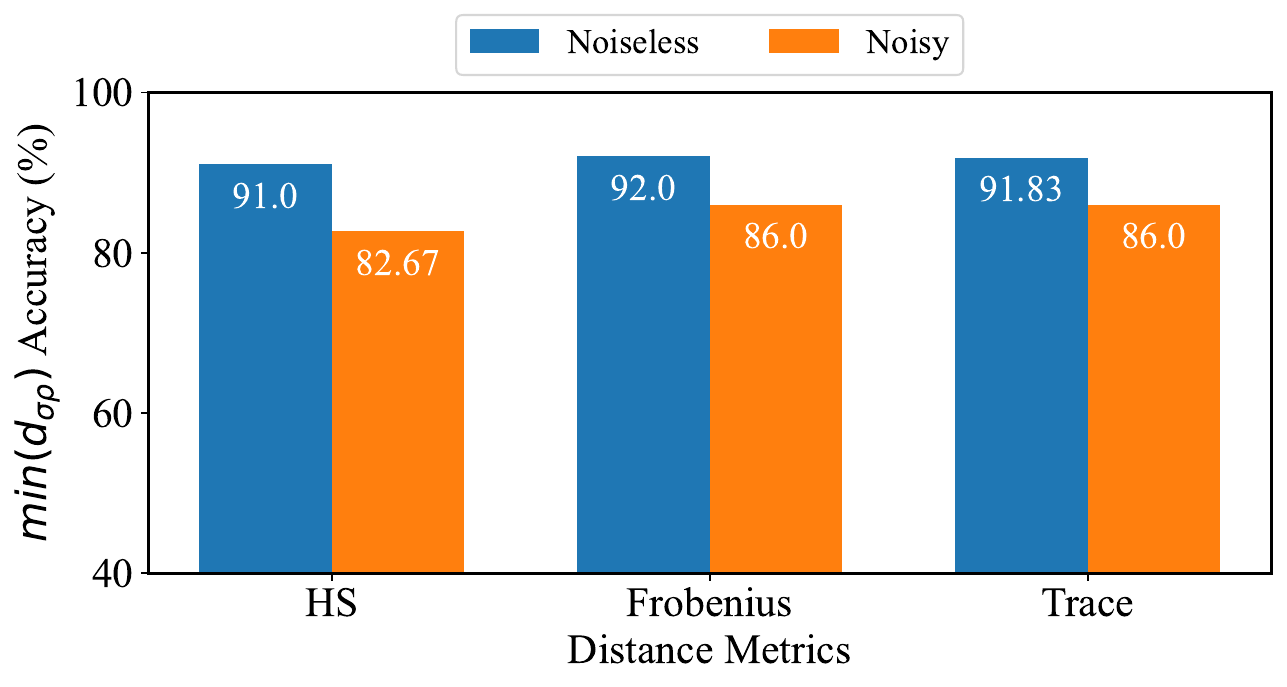}
        \vspace{-3mm}
        \caption{Figure illustrating the accuracy of labels generated based on $\min$ intra-class state distances compared to the true labels for the reduced MNIST-10 dataset with a latent dimension of $d = 8$, evaluated under both noiseless and noisy environments.}
        \label{mnist_10_validation}
        \vspace{-2mm}
\end{figure}

\subsection{Intra-Class Encoder State Similarity (ESS)} \label{sec:ess}
In quantum machine learning models, classical data is mapped into the quantum Hilbert space using an encoding circuit ($\phi$). A variety of encoding techniques exist, but the main purpose of all of them is to embed the classical data in a way that leads to minimal loss of information, i.e., the separability of the encoded states, represented by the density matrix $\rho$, belonging to the same class (intra-class) is low, while that between different classes (inter-class) is high. Theoretically, we can say that given any input space $\mathcal{X}$, corresponding labels $\mathcal{Y}$, an encoding circuit $\phi : \mathcal{X} \to \mathcal{D(H)}$ that maps classical data to density matrices in the quantum Hilbert space $\mathcal{H}$, and a distance metric $d(\cdot, \cdot)$, the quantum encoding circuit $\phi$ preserves the class structure in the sense that:
\begin{align*}
    d(\phi(x_1), \phi(x_2)) &< d(\phi(x_1), \phi(x_3)) \\
    d(\phi(x_1), \phi(x_2)) &< d(\phi(x_2), \phi(x_3)),
\end{align*}
where $x_1, x_2$ belong to the same class, and $x_3$ belongs to a different class. Thus, for any input data $x$, we can easily estimate the corresponding label $\in \mathcal{Y}$ by calculating the matrix distance between encoded state and all the other encoded states as, $d(\phi(x), \phi(x_i)) : x_i \in \mathcal{X}$, average them based on the class and assign it the label corresponding to the class with $\min d$. 

We take an empirical approach to validate intra-class state similarity by assigning the encoded state's density matrix the label of its nearest class in the quantum Hilbert space $\mathcal{H}$ and verifying how accurately it matches the true label. We conducted experiments using the angle encoding technique (with $R_Z$ and $R_X$ gates) across different datasets and employing various distance metrics. Specifically, given two density matrices, $\sigma$ and $\rho$, we use the following Frobenius norm, trace norm and a normalized distance metric derived from the Hilbert-Schmidt inner product \( D_{\text{HS}} = 1 - \frac{|\text{Tr}(\sigma^\dagger \rho)|}{\text{dim}(\sigma)}\) to calculate the distance between the two states in $\mathcal{H}$. 
% \begin{align*} 
%     D_{\text{HS}}(\sigma, \rho) &= 1 - \frac{|\text{Tr}(\sigma^\dagger \rho)|}{\text{dim}(\sigma)} \\ 
%     D_{\text{Frobenius}}(\sigma, \rho) &= \sqrt{\text{Tr}[(\sigma - \rho)^2]} \\
%     D_{\text{Trace}}(\sigma, \rho) &= \frac{1}{2} \text{Tr}\left[\sqrt{(\sigma - \rho)^\dagger(\sigma - \rho)}\right] 
% \end{align*}
We avoided fidelity-based distance metrics like the Hellinger distance due to it's high computation time (almost $>20\times$ compared to the Frobenius norm) while offering no performance improvement. 

\begin{table}[!t]
    \vspace{-2mm}
    \centering
    \caption{Performance comparison of label assignment based on $\min$ intra-class distance with respect to true labels, across various 4-class datasets and distance metrics. Frobenius norm appears to be the most optimal choice due to its superior performance and efficient computational cost.}
    \label{tab_intra_class_similarity}
    \vspace{-1mm}
    \begin{tabular}{cccccc}
    \cmidrule(lr){2-5}
    \multicolumn{1}{c}{} & \multicolumn{4}{c}{$\min(d_{\sigma\rho})$ Accuracy} & \multicolumn{1}{c}{}\\
    \cmidrule(lr){1-6}
    Metrics    & MNIST & Fashion & Kuzushiji & Letters &  Time (s) \\
    \cmidrule(lr){1-6}
    HS              & 90.0  & 89.6          & 69.3      &  79.0     & 7.4         \\
    Frobenius       & 91.7  & \textbf{90.7}          & \textbf{75.0}      &  \textbf{86.3}     & \textbf{2.2}   \\ 
    Trace           & \textbf{92.0}  & 90.3          & 74.0      &  86.0     & 9.8     \\
    \bottomrule
    \end{tabular}
    \vspace{-3mm}
\end{table}

Fig.~\ref{mnist_10_validation} illustrates how accurately we can identify classes in a portion (50\%) of a reduced MNIST-10 dataset using intra-class state similarity. Under both noiseless and noisy scenarios, the Frobenius norm outperforms the HS and Trace norms in determining the correct classes. Table~\ref{tab_intra_class_similarity} further presents the performance of different 4-class datasets under noisy conditions. Here too, we observe that, on average, using the Frobenius norm to calculate the distance between density matrices results in better performance while requiring considerably less computation time, making it the optimal choice of distance metric. Furthermore, we believe ESS can be used to estimate the performance of encoding methods and to estimate the separability of datasets when mapped to $\mathcal{H}$ without the need for training the entire QNN from scratch.

%As discussed earlier, unlike classical neural networks where a batch of input data can be processed in a single iteration on a GPU, QML circuits can process only one input sample at a time. For each input data point $X$, the sample is first encoded into the QML circuit and then sent to the quantum cloud for execution. The cloud returns the measured outputs for that specific input sample, which may undergo further local post-processing, such as applying a softmax function. We take advantage of this by intermittently introducing poisoned samples into the dataset to obfuscate the original data, thereby degrading the performance of the adversarial QML model when trained on this altered dataset. However, the user locally disregards the outputs of the QML models executed with the poisoned samples, preserving the model's overall performance. Consequently, the user can reduce the value of the dataset at an increased execution cost.

\subsection{QUID Poisoning Strategy}
The goal of an adversary in data poisoning attacks is to create a poisoning dataset $\mathcal{D}_p \subset \mathcal{D}_{tr}$ which leads to maximum performance degradation of QNN ($\mathcal{F}$) on $\mathcal{D}_{test}$ or,
\begin{align*}
    \max \text{Loss}\;\mathcal{F}(\theta;\mathcal{D}_{test})
\end{align*}
where $\theta$ represents the set of parameters of $\mathcal{F}$. 
%Given, that the adversary does not directly have access to the parameters of the QNN, is it possible to determine the optimal poisoned samples which lead to maximum performance degradation? 
We use the intra-class quantum state similarity concept to find adversarial labels which leads to high performance degradation. Given that for any dataset, intra-class distances and inter-class distances in $\mathcal{H}$ should be low and high respectively, we modify the labels of a subset ($\varepsilon$) of the training data $\mathcal{D}_{tr}$ which leads to high intra-class distances. Specifically, given the training dataset $\mathcal{D}_{tr}$, we first split it into clean $\mathcal{D}_c$ and poisoned dataset $\mathcal{D}_{p}$ based on the poison ratio $\varepsilon$. For samples $\{ x_i \mid (x_i, y_i) \in \mathcal{D}_p \}$, find density matrix of the encoded states $\phi(x_i) : x_i \in \mathcal{D}_p$, calculate distance between the encoded states $\phi(x_i) : x_i \in \mathcal{D}_c$ and assign the class label with $\mathbf{\max}$ distance. An algorithmic description of QUID's label poisoning technique is described in Algorithm \ref{alg:qnn_poison}. Essentially, we propose a greedy algorithm to determine optimal labels, in contrast to \cite{paudice2019label} which relies on a brute force training method involving iterating through all possible labels for each sample and selecting the label that results in the maximum loss.

\begin{algorithm}[!t]
\caption{QUID's Label Poisoning Procedure}
\label{alg:qnn_poison}
    \begin{algorithmic}[1]
    \REQUIRE Training data $\mathcal{D}_{tr} = {(x_i, y_i)}_{i=1}^n$, Poison ratio $\varepsilon$, Encoding circuit $\phi$, Distance metric $d(\cdot,\cdot)$ for density matrices
    \ENSURE Poisoned dataset with modified labels
        \STATE Split $\mathcal{D}_{tr}$ into clean set $\mathcal{D}_c$ and poison set $\mathcal{D}_p$ with ratio $\varepsilon$
        \STATE $\mathcal{C} \gets \text{unique}(y)$ \COMMENT{Set of unique classes}
        \STATE $\rho_c \gets {\phi(x) : (x,y) \in \mathcal{D}_c}$ \COMMENT{Encoded clean states}
        \STATE $\rho_p \gets {\phi(x) : (x,y) \in \mathcal{D}_p}$ \COMMENT{Encoded poison states}
        \FOR{$\rho_i \in \rho_p$}
        \STATE $D_{cls} \gets {}$ \COMMENT{Dictionary for class-wise distances}
        \FOR{$c \in \mathcal{C}$}
        \STATE $\rho_c^{(c)} \gets {\rho : \rho \in \rho_c, y = c}$ \COMMENT{States of class $c$}
        \STATE $D_{cls}[c] \gets \frac{1}{|\rho_c^{(c)}|}\sum_{\rho \in \rho_c^{(c)}} d(\rho_i, \rho)$
        \ENDFOR
        \STATE $y_i^{new} \gets \mathrm{arg}\max \limits_{c \in \mathcal{C}} D_{cls}[c]$ \COMMENT{Assign class with max dist.}
    \ENDFOR
    \RETURN $\mathcal{D}_c \cup {(x_i, y_i^{new})}:{i \in \mathcal{D}_p}$
    \end{algorithmic}
\end{algorithm}
\vspace{-2mm}

\section{Evaluation}
\subsection{Experimental Setup} \label{setup}

\noindent \textbf{Device:} We conducted all noiseless experiments on Pennylane's \cite{bergholm2018pennylane} \textit{lightning.qubit} device using the ``Adjoint'' differentiation method to calculate gradients. For noisy training, we employed the \textit{default.mixed} device and introduced \textit{Amplitude Damping} and \textit{Depolarizing Channel} noise after each gate in the circuit, with a high error probability of $p = 0.05$ (unless stated otherwise) to effectively simulate the noisy nature of current hardware. Considering that generic methods for calculating gradients, such as backpropagation, are not viable on actual quantum devices, we used the Simultaneous Perturbation Stochastic Approximation (SPSA) method for gradient calculation in our QNN training. This approach generates noisy gradients, providing a more realistic representation of the performance expected from real quantum hardware and ensuring that our experimental results align more closely with practical quantum computing environments.

\noindent \textbf{Training:} For evaluation, we used PQC-1, 6 and 8 to build our QNNs (PQC-$x$ represents circuit-$x$ in \cite{sim2019expressibility}), initialized with random weights. In most of the experiments we used a 4-qubit or 8-qubit QNN and angle encoding to encode classical features, consistent with recent QML works \cite{wang2022quantumnas, anagolum2024elivagar}. For training the QNNs we used CrossEntropyLoss() as our classification loss function. The hyperparameters used for training are; Epochs: 30, learning\_rate ($\eta$): 0.01, batch\_size = 32 and optimizer: Adam. All training are done on an Intel Core i9-13900K CPU with 64GB of RAM.

\noindent \textbf{Dataset:} Since NISQ devices struggle with large images due to limited qubits, high error rates and complex data encoding, similar to recent works \cite{wang2022qoc, anagolum2024elivagar}, we conduct all experiments using a reduced feature set of MNIST, Fashion, Kuzushiji and Letters datasets with latent dimension $d = 8$ (from original 28$\times$28 image) generated using a convolutional autoencoder \cite{alam2021quantum}. Thus, for each dataset, we create a smaller 4-class dataset from these reduced feature sets i.e., MNIST-4 (class 0, 1, 2, 3), Fashion-4 (class 6, 7, 8, 9), Kuzushiji-4 (class 3, 5, 6, 9) and Letters-4 (class 1, 2, 3, 4) with each having 1000 samples (700 for training and 300 for testing). Since each of these datasets is of dimension $d = 8$, we encode 2 features per qubit. Number of shots/trials is set to 1000 for all experiments.

\begin{figure}[!t]
        \vspace{-2mm}
        \centering 
        \includegraphics[width=\linewidth]{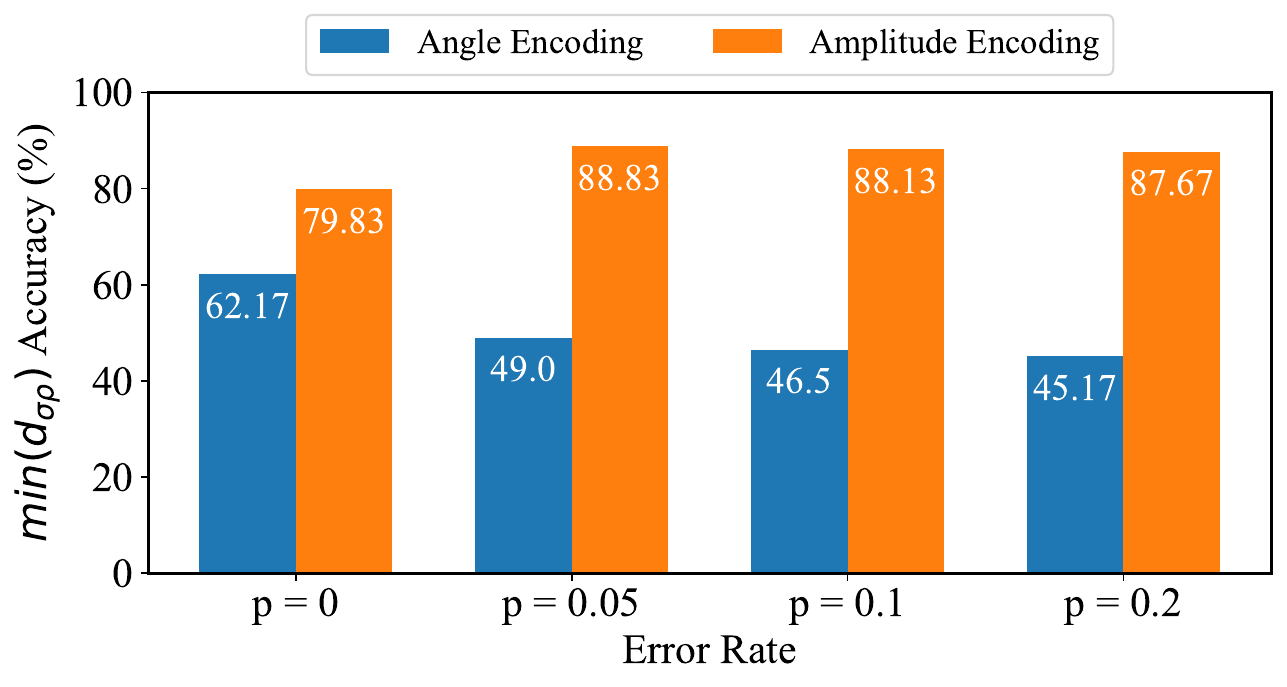}
        \vspace{-6mm}
        \caption{Performance of angle encoding (4 features/qubit) and amplitude encoding when using ESS to determine embedding circuit performance under varying noise levels.}
        \label{fig:angle_amplitude}
        \vspace{-2mm}
\end{figure}

\begin{table}[!b]
    \vspace{-1mm}
    \centering
    \caption{Test accuracies for label-flip (LF) and bi-level poisoning attacks on different datasets, where poisoned samples (PS) $x_i \in \mathcal{D}_p$ are randomly generated. QUID achieves maximal performance degradation when only the labels are poisoned.}
    \label{tab:lf_bilevel}
    \vspace{-2mm}
    \begin{tabular}{cccccc}
    \cmidrule(lr){3-6}
    \multicolumn{2}{c}{} & \multicolumn{2}{c}{Label-flip (LF)} & \multicolumn{2}{c}{Bi-level (LF + PS)}\\
    \cmidrule(lr){1-6}
    Dataset    & Baseline & Random & QUID & Random & QUID  \\
    \cmidrule(lr){1-6}
    MNIST-4              & 87.3  & 76.7          & \textbf{7.7}      &  77.3 &  10.3              \\
    Fashion-4       & 82.3  & 75.3          & \textbf{1.33}      &  75.0  &  47.6      \\
    Kuzushiji-4       & 73.6  & 65.6          & \textbf{24.3}      &  70.7 &  28.9       \\ 
    Letters-4           & 76.0  & 60.6          & \textbf{5.7}      &  74.7 &  18.3        \\
    \bottomrule
    \end{tabular}
    \vspace{-2mm}
\end{table}

\subsection{Analysis} \label{sec:results}
\noindent \textbf{Angle vs. Amplitude Encoding:} To evaluate how effectively the proposed ESS (Section~\ref{sec:ess}) can be used to measure the performance of state preparation circuits on any given dataset, we compare the intra-class state similarity for angle and amplitude encoding under varying noise levels ($p$: error probability) on a reduced MNIST-10 dataset with a latent dimension of 16. Considering that amplitude encoding can encode $2^n$ features in $n$ qubits, for fairness, we used a 4-qubit circuit for angle encoding, with each qubit having four rotation gates ($R_Z$, $R_X$, $R_Z$, $R_X$), to provide a better comparison against amplitude encoding. Fig.~\ref{fig:angle_amplitude} shows the performance comparison of both encoding techniques. The results are interesting, as amplitude encoding performs worse under noiseless conditions. However, it is evident that amplitude encoding outperforms angle encoding when encoding larger datasets, likely due to the loss of information caused by applying a high number of sequential rotation gates in angle encoding which is further amplified by noise. Conversely, angle encoding performs well for smaller datasets, as demonstrated in Fig.~\ref{mnist_10_validation}, where two features are encoded per qubit.

\noindent \textbf{Label-flip vs. Bi-level:} Even though QUID is primarily a label-flipping (LF) technique, we compare its performance against random bi-level poisoning. In this method, apart from flipping labels, we also replace samples of the poisoned data ($\mathcal{D}_p$) i.e.  $\{x_i \mid x_i \in \mathcal{D}_{p} \subset \mathcal{D}_{tr}\}$ ($\varepsilon = 0.5$), with random features and calculate the optimal poisoned labels for these random features. Table~\ref{tab:lf_bilevel} shows the performance comparison of label-flipping versus bi-level poisoning when training a 4-qubit QNN on various datasets under noise with error rate $p = 0.1$. Our experiments indicate that, in most cases, adding randomly generated poisoned samples (PS) actually leads to better performance (i.e. less accuracy \& loss degradation) compared to the scenario of using only LF with QUID. For the remaining analysis, we use only the label-flipping technique.

\begin{table}[!t]
    \vspace{-2mm}
    \centering
    \caption{Test accuracies for QUID when trained on noiseless and noisy environment on different datasets  with $\varepsilon = 0.5$. QUID performs better than random label flipping under noisy conditions, demonstrating it's robustness to noise.}
    \label{tab:noisy_noiseless}
    \vspace{-2mm}
    \begin{tabular}{cccccc}
    \cmidrule(lr){2-6}
    \multicolumn{1}{c}{} & \multicolumn{3}{c}{Noiseless} & \multicolumn{2}{c}{Noisy ($p = 0.05$)}\\
    \cmidrule(lr){1-6}
    Dataset    & Baseline & Random & QUID & Random & QUID  \\
    \cmidrule(lr){1-6}
    MNIST-4              & 94.3  & 91.6          & 43.9      &  89.9 &  \textbf{31.6}  \\
    Fashion-4       & 85.0  & 83.3          & 31.9      &  82.9  &  \textbf{7.9}      \\
    Kuzushiji-4       & 81.3  & 74.6          & 37.0      &  72.0 &  \textbf{27.3}       \\ 
    Letters-4           & 88.9  & 77.9          & 30.0      &  75.6 &  \textbf{12.6}        \\
    \bottomrule
    \end{tabular}
    \vspace{-2mm}
\end{table}

\begin{table}[!b]
    \vspace{-1mm}
    \centering
    \caption{Test accuracies of QUID with poison ratio $\varepsilon = 0.5$ on real hardware noise model.}
    \label{tab:real_noise}
    \vspace{-2mm}
    \begin{tabular}{cccccc}
    \cmidrule(lr){2-6}
    \multicolumn{1}{c}{} & \multicolumn{3}{c}{IBM\_Brisbane} & \multicolumn{2}{c}{IBM\_Kyiv}\\
    \cmidrule(lr){1-6}
    Dataset    & Baseline & Random & QUID & Random & QUID  \\
    \cmidrule(lr){1-6}
    MNIST-4              & 83.1  & 71.7          & \textbf{10.3}      &  73.6 &  \textbf{8.9}              \\
    Fashion-4       & 80.7  & 73.3          & \textbf{8.7}      &  73.8  &  \textbf{7.6}      \\
    \bottomrule
    \end{tabular}
    \vspace{-2mm}
\end{table}

\noindent \textbf{Noiseless vs. Noisy Simulations:} Table~\ref{tab:noisy_noiseless} presents the performance comparison of QUID when training a QNN under both noiseless and noisy conditions. From the table, it is evident that QUID demonstrates significantly better performance under noisy conditions compared to noiseless conditions. Specifically, under noise, QUID leads to an additional accuracy degradation of up to 24\%, whereas random flipping causes only about 2\% degradation. This indicates that even under noise, QUID remains effective and demonstrates a significant advantage over classical approaches.
%For example, at a poison ratio of $\varepsilon = 0.5$, QUID reduces performance by $78\%$ under noisy conditions compared to 53\% under noiseless conditions on Fashion-4 dataset. This result highlights the robustness of QUID against noise.

\noindent \textbf{Real Hardware Noise:} Given the limited availability of real quantum devices, we directly integrated noise models from actual quantum hardware. Specifically, we employed the noise models of IBM\_Kyiv and IBM\_Brisbane, and incorporated them into the \textit{qiskit.aer} device to simulate the noise characteristics of these devices. Table~\ref{tab:real_noise} lists the performance of QUID and random label-flipping on real hardware noises. We observe that even in this scenario, QUID is able to significantly degrade the performance of the baseline model by $\approx73\%$ on IBM\_Brisbane's noise and $\approx72\%$ on IBM\_Kyiv's noise.

\begin{table*}[!t]
    \vspace{-4mm}
    \centering
    \caption{Comparison of test accuracy after training QNNs with various PQC architectures on different poison ratios ($\varepsilon$) under noise with error rate $p = 0.05$. Cplx. refers to the complexity of the PQC in terms of circuit gate count and depth.}
    \label{tab:poison_ratio}
    \vspace{-2mm}
    \begin{tabular}{ccccccccccc}
    \cmidrule(lr){3-11}
    \multicolumn{2}{c}{} & \multicolumn{1}{c}{$\varepsilon = 0$} & \multicolumn{2}{c}{$\varepsilon = 0.1$} & \multicolumn{2}{c}{$\varepsilon = 0.3$} & \multicolumn{2}{c}{$\varepsilon = 0.5$} & \multicolumn{2}{c}{$\varepsilon = 0.7$} \\
    \cmidrule(lr){1-11}
    Datasets    & PQC (cplx.) & Baseline & Random & QUID & Random &  QUID & Random &  QUID & Random &  QUID \\
    \cmidrule(lr){1-11}
    \multirow{3}{*}{MNIST-4}              & PQC-1 (low) & 93.3  & 93.0          & 92.6      &  92.3     & 82.3 &  89.9     & 31.6 &  53.3     & 0.99        \\
    & PQC-8 (mid) & 81.3  & 80.6          & 77.3      &  77.6     & 69.3 &  74.6     & 20.9 &  40.6     & 5.63        \\
    & PQC-6 (high) & 72.6  & 54.3          & 46.9      &  43.3     & 21.3 &  52.3     & 20.9 &  33.6     & 20.9        \\
    \cmidrule(lr){1-11}
    Fashion-4       & PQC-1 & 87.0  & 86.3          & 82.9      &  87.0     & 76.3 &  82.9     & 7.9 &  75.3     & 0.66   \\ 
    \cmidrule(lr){1-11}
    Kuzushiji-4           & PQC-1 & 77.6  & 76.3          & 27.3      &  76.9     & 70.6 &  72.0     & 27.3 &  35.3     & 0.66    \\
    \cmidrule(lr){1-11}
    Letters-4           & PQC-1 & 84.3  & 83.3          & 81.6      &  81.3     & 61.3 &  75.6     & 12.6 &  38.9     & 1.99     \\
    \bottomrule
    \end{tabular}
    \vspace{-1mm}
\end{table*}

\noindent \textbf{Poison Ratio ($\boldsymbol{\varepsilon}$):} We also evaluate the performance of QUID using different poison ratios, which helps determine the threshold beyond which the model's performance decreases drastically. All training in this evaluation is conducted under noise ($p = 0.05$) using the SPSA gradient calculation method. Table~\ref{tab:poison_ratio} lists the test accuracies of various QNNs trained across different datasets and poison ratios ($\varepsilon$). The main reason for selecting PQC-1, PQC-6, and PQC-8 is to demonstrate how differences in expressivity and noise across various PQC architectures impact the effectiveness of QUID. PQC-1 is the least expressive, while PQC-6 is the most expressive \cite{sim2019expressibility}, but with higher gate count and depth, leading to greater noise accumulation. As shown in Table~\ref{tab:poison_ratio}, this results in QUID being less effective in PQC-6, especially for higher epsilon values, as noise eventually randomizes the outputs. Conversely, for PQC-1, the QUID technique remains highly effective, with accuracy dropping to sub-1\%. 
%As expected, we observe from the table that baseline models of more complex PQCs like PQC-6 perform worse compared to PQC-8 under hardware noise, while the circuit with the least gate count and depth significantly outperforms both. 
Consequently, we selected PQC-1 for noisy training on the other datasets. 

Another important observation is that with a small poison ratio ($\varepsilon \leq 0.1$), there appears to be minimal performance degradation. However, for $\varepsilon > 0.5$, the performance suffers diminishing returns, as the trained model on QUID achieves a test accuracy less than $20\%$, worse than random prediction ($25\%$ for 4-class dataset), rendering the model completely unusable for the victim. In all cases, the proposed QUID technique performs significantly better than the random label-flipping technique.

\begin{table}[!b]
    \vspace{-3mm}
    \centering
    \caption{Test accuracies for QUID when trained w/ and w/o SS-DPA \cite{levine2020deep}, a SOTA defense against label-flipping attacks.}
    \label{tab:defense}
    \vspace{-2mm}
    \begin{tabular}{cccccc}
    \cmidrule(lr){3-6}
    \multicolumn{2}{c}{} & \multicolumn{2}{c}{$\varepsilon = 0.3$} & \multicolumn{2}{c}{$\varepsilon = 0.5$}\\
    \cmidrule(lr){1-6}
    Dataset    & Baseline & w/o Def. & SS-DPA & w/o Def. & SS-DPA \\
    \cmidrule(lr){1-6}
    MNIST-4              & 94.3  & 91.0          & 91.3      &  43.9 & 49.3  \\
    Fashion-4       & 85.0  & 75.0          & 75.3      &  31.9  & 41.0       \\
    Kuzushiji-4       & 81.3  & 73.7          & 80.3      &  37.0 &  40.7    \\ 
    Letters-4           & 88.9  & 72.0         & 76.7      &  30.0 &  21.0      \\
    \bottomrule
    \end{tabular}
    \vspace{0mm}
\end{table}

\noindent \textbf{Defense Evaluation:} %\hl{New!} 
Although label-flipping attacks are relatively easy to defend against in the classical domain using data sanitization techniques \cite{taheri2020defending, steinhardt2017certified}, implementing such defenses in the current quantum cloud environment poses significant challenges, as the victim cannot intervene once the quantum circuit and training data ($\mathcal{D}_{tr}$) are submitted for training. Therefore, we evaluate the effectiveness of QUID against a practical training-based defense technique, SS-DPA \cite{levine2020deep}, which is specifically designed to mitigate label-flipping attacks. SS-DPA leverages subset aggregation and ensemble learning together with semi-supervised learning. Table~\ref{tab:defense} illustrates the performance of QUID with and without SS-DPA, trained in a noiseless setting. The hyperparameter $k$ in SS-DPA is set to 3 due to the smaller dataset sizes. While the application of SS-DPA improves performance by a small amount in most cases, it is not very effective for large $\varepsilon$. Moreover, it is extremely expensive in terms of computation overhead, as it requires the training and retraining of multiple QNNs. Future research should focus on the development of effective and efficient defense techniques tailored specifically to QMLs.

%Furthermore, while state-of-the-art (SOTA) bagging-based defenses \cite{jia2021intrinsic, wang2022improved} can effectively counter QUID, they require training multiple QNNs in parallel. This approach is infeasible due to the extremely high cost of current quantum hardware. Future research should focus on developing efficient defense techniques tailored to QMLs.

\section{Discussions}

\begin{figure}[!t]
        \vspace{0mm}
        \centering 
        \includegraphics[width=\linewidth]{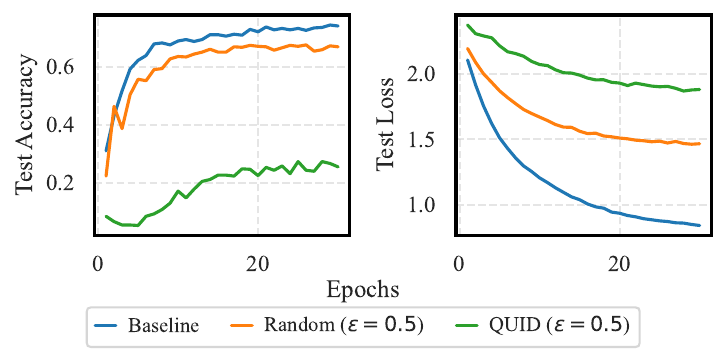}
        \vspace{-6mm}
        \caption{Test performance of an 8-qubit, 2-layer QNN on the reduced MNIST-10 dataset with $\varepsilon = 0.5$. QUID significantly degrades model performance on $\mathcal{D}_{test}$, even when trained on larger datasets.}
        \label{fig:scalability_8q_mnist_10}
        \vspace{-5mm}
\end{figure}

\noindent \textbf{Scalability:} To demonstrate the effectiveness of QUID for larger QNNs trained on larger datasets, we designed an 8-qubit QNN with two layers of PQC-1 \cite{sim2019expressibility}. We trained this model on the MNIST-10 dataset, reduced to a latent dimension of $d = 16$, using 2000 samples with a 70:30 train-test split. Fig.~\ref{fig:scalability_8q_mnist_10} shows the test loss and test accuracy of the victim model when trained on a poisoned dataset with a poison ratio $\varepsilon = 0.5$, using both random label flipping and QUID's label poisoning technique. The results clearly illustrate that QUID outperforms random label flipping techniques, even for in this scenario. Specifically, QUID results in additional $\approx41\%$ accuracy degradation and incurs $\approx0.4$ higher test loss compared to random flipping, further validating its effectiveness.

\noindent \textbf{Applications:} QUID is not only effective against indiscriminate poisoning attacks but can also be adapted for targeted attacks, where the adversary selectively flips labels of a specific class. This tactic allows the QNN to perform well on most test data while deliberately misclassifying the targeted class, serving the adversary's objective. Additionally, QUID can also be used for secure dataset releases, as demonstrated in \cite{fowl2021adversarial}. By injecting poisoned samples with labels determined by QUID, legitimate users can prevent unauthorized usage. Attackers unknowingly training on such datasets would produce suboptimal models, rendering their efforts ineffective.

\noindent \textbf{Limitations:} 
The primary limitation of QUID lies in its reliance on the density matrix to assign optimal labels. While density matrices can be reconstructed on real hardware using techniques such as quantum state tomography, the measurement cost scales exponentially with the number of qubits. However, recent works \cite{ahmed2021quantum, lange2023adaptive} show that classical models can now approximate these matrices accurately. This suggests that QUID can remain practical and scalable even when applied to larger QML models. Other relevant works \cite{chen2021low} have also proposed more efficient simulation techniques. Although expectation-value-based measurement techniques are computationally cheaper, they come at the cost of significant information loss, which limits the effectiveness of QUID.

%Furthermore, although data sanitization techniques \cite{taheri2020defending, steinhardt2017certified} have been largely successful in the classical domain for mitigating performance degradation caused by label-flipping attacks, such defenses prove ineffective in the current quantum cloud environment, as the victim cannot intervene once the quantum circuit and training data ($\mathcal{D}_{tr}$) have been submitted to the cloud for training. SOTA bagging-based defenses \cite{jia2021intrinsic, wang2022improved} could effectively defend against QUID but require the high cost of training multiple QNN classifiers.

% Another drawback of QUID is that, as a predominantly label-flipping attack, it has been found to be relatively easy to defend against in the classical domain. For instance, data sanitization techniques \cite{taheri2020defending, steinhardt2017certified} have proven effective in detecting poisoned samples. However, implementing such defenses in the current quantum cloud environment is challenging, as the victim cannot intervene once the quantum circuit and training data ($\mathcal{D}_{tr}$) are submitted for training. Furthermore, while state-of-the-art (SOTA) bagging-based defenses \cite{jia2021intrinsic, wang2022improved} can effectively counter QUID, they require training multiple QNNs in parallel. This approach is infeasible due to the extremely high cost of current quantum hardware. Future research should focus on developing efficient defense techniques tailored to QMLs.

\section{Conclusion}
In recent years, there has been significant research evaluating the security of quantum computers in cloud settings. In this work, for the first time, we focus on one such threat to QML models, specifically, data poisoning attacks. An adversary in the cloud can significantly manipulate the performance of QNN models by tampering with the dataset. To this end, we propose a novel label poisoning attack, QUID, based on intra-class encoder state similarity (ESS), which can degrade the performance of QML models. Our extensive experiments under noise reveal that QUID significantly outperforms random poisoning, degrading QNN accuracy by more than $90\%$ and by up to $50\%$ even when classical defense techniques are applied. Thus, future work should focus on developing robust defense techniques to effectively mitigate this threat in the context of QML.

\section*{Acknowledgment}
The work is supported in parts by NSF (CNS-2129675, CCF-2210963, OIA-2040667, DGE-1821766 and DGE-2113839), gifts from Intel and IBM Quantum Credits.

\bibliographystyle{ACM-Reference-Format}
\bibliography{refs}

\end{document}